\documentclass[
twocolumn,
superscriptaddress,
amsmath,amssymb,
aps]{revtex4-2}

\usepackage{graphicx}
\usepackage{bm}
\usepackage{hyperref}
\usepackage{color} 
\usepackage{txfonts}
\usepackage[dvipsnames]{xcolor}
\usepackage{ragged2e}
\usepackage{amsmath}
\usepackage{ulem}

\newcommand {\fabsq}[1] {\left| #1 \right|^{2}}

\newcommand{\braket}[2]{\langle#1|#2\rangle}

\newcommand {\cE} {{\cal E}}
\newcommand {\cH} {{\cal H}}
\newcommand {\XX} {{\chi}}
\newcommand {\PP} {{\cal P}}

\begin{document}

\title{Piston control in a two-ion quantum device}

\author{Jing Li}
\affiliation{School of Physical Science and Technology, Nantong University, Nantong 226001, Jiangsu, China}
\affiliation{School of Physics, University College Cork, T12 K8AF Cork, Ireland}

\author{E. Ya. Sherman}%
\affiliation{Department of Physical Chemistry, University of the Basque Country UPV/EHU, 48940 Leioa, Spain}%
\affiliation{IKERBASQUE, Basque Foundation for Science, 48009 Bilbao, Spain}
\affiliation{EHU Quantum Center, University of the Basque Country UPV/EHU, 48940 Leioa, Spain}

\author{Andreas Ruschhaupt}
\affiliation{School of Physics, University College Cork, T12 K8AF Cork, Ireland}

\begin{abstract}
We propose a scheme for piston control in a two-ion quantum device with motion confined to orthogonal axes. In this system, one ion plays the role of a “classical” piston driven by the Coulomb interaction with the other ion, whose quantum motion is controlled through modulation of its trapping potential. The stationary state is determined self-consistently, taking quantum effects into account. We identify a narrow quantum regime of the ground state connecting two broad classical regimes. We further design inverse-engineering protocols to control the motion of the “classical” ion. 
The proposed control scheme provides a useful route toward controlled piston dynamics in microscopic quantum devices.
\end{abstract}

\maketitle

\section{Introduction}
\label{sec:1}

In classical heat engines, the working medium typically contains a macroscopic number of particles, often far exceeding the Avogadro number, and its behavior is governed by thermodynamic laws and equations of state \cite{callen_textbook}. In contrast, small systems composed of only a few particles operate in regimes where the extracted work is small and quantum effects become important \cite{Kosloff13,Goold_2016,Myers2022}. In such systems, fluctuations and coherence play an essential role, and thermodynamic processes are governed by the underlying quantum dynamics \cite{yamazaki2025TLL,du2022,Francica2017,Poschinger2019}. In particular, the behavior of small working media and their coupling to controllable degrees of freedom often has to be analyzed at the level of the microscopic model.

This has motivated the study of quantum heat engines in a variety of controllable platforms, including quantum dots \cite{Sothmann_2014,Josefsson2018}, trapped ions \cite{Lutz2012,Kilian2016,carvalho2026}, ultracold atoms \cite{brantut2013,Widera21,Koch2022,Jing22} and optomechanical systems \cite{Pierre2014,Mari_2015}. Among them, trapped ions and ultracold atoms are particularly attractive because both the confining potentials and the interparticle couplings can often be manipulated with high precision. Many-body working media, such as Bose-Einstein condensates, further enrich this setting through tunable interactions and collective dynamics \cite{Charalambous_2019,Mininni2024}. Since microscopic engines necessarily operate in finite time, quantum control becomes a central ingredient, and techniques such as shortcuts to adiabaticity \cite{STAreview,Campo2014,Palmero2014,Palmero2015,Whitty2020} and optimal control provide natural tools for designing efficient driving protocols \cite{Glaser2015,Steve2025}.

A key element of most heat engines is the piston, which converts internal energy into mechanical motion and hence performs work. While this concept is straightforward in macroscopic systems, its realization at the microscopic scale is less obvious. In particular, in a small interacting quantum system it is not always clear which degree of freedom should play the role of the piston, how it should be coupled to the working medium, and how its motion can be controlled in a physically transparent way. Piston-like degrees of freedom have been discussed, for example, in optomechanical systems \cite{Mari_2015}, controlled trapped-ion motion experiments \cite{Bowler2012,Poschinger2012}, and interacting cold atoms \cite{Mossman2018,Jing_2024,rodin2024piston}. These examples show that microscopic piston motion can be realized in different settings, while also highlighting the usefulness of simple models in which the roles of working medium and piston are clearly separated.

In this paper, we consider a system of two ions confined in orthogonal harmonic traps and interacting via the Coulomb force taking advantage of the geometry, see Fig. \ref{fig:system}. 
The lighter ion is treated quantum mechanically and serves as the working medium, while the heavier ion, due to its large mass, can be approximated as a classical particle and acts as a piston. 
The coupling leads to a hybrid quantum--classical dynamics with backaction: the position of the heavy "classical" ion enters the quantum Hamiltonian, while the force acting on it is determined by the quantum state of the light ion. In this way, a piston-like degree of freedom arises naturally from the interaction rather than being introduced externally.

We will see that the interplay between harmonic confinement and Coulomb interaction leads to distinct regimes of the ground state of the system, including a narrow region where quantum effects become essential. At the same time, the same coupling makes it possible to control the motion of the classical ion through quantum manipulation of the lighter one: by modulating the trapping frequency of the light particle, the position of the heavy ion can be steered. We design the control function by inverse engineering, in which a desired piston trajectory is specified and the corresponding time dependence of the trapping frequency is obtained self-consistently by inverse engineering. The system therefore provides a minimal platform combining a clear spatial separation between working medium and piston, self-consistent backaction, and the possibility of high-fidelity control.

In Sec.~\ref{sec:model}, we present the full quantum description of the system and then derive a hybrid quantum-classical approximation. In Sec.~\ref{sec:stationary}, we analyze the stationary states and identify three regimes of the ground state, for which complementary approaches are developed and compared. We then turn to the time-dependent problem in Sec.~\ref{sec:dynamics}, where we investigate the control of the heavy ion through modulation of the trapping frequency of the light ion and derive the corresponding control scheme. Finally, we summarize our results in the Conclusion.

\section{System, model and approximations}
\label{sec:model}

\begin{figure}[t]
\includegraphics[width=0.86\linewidth]{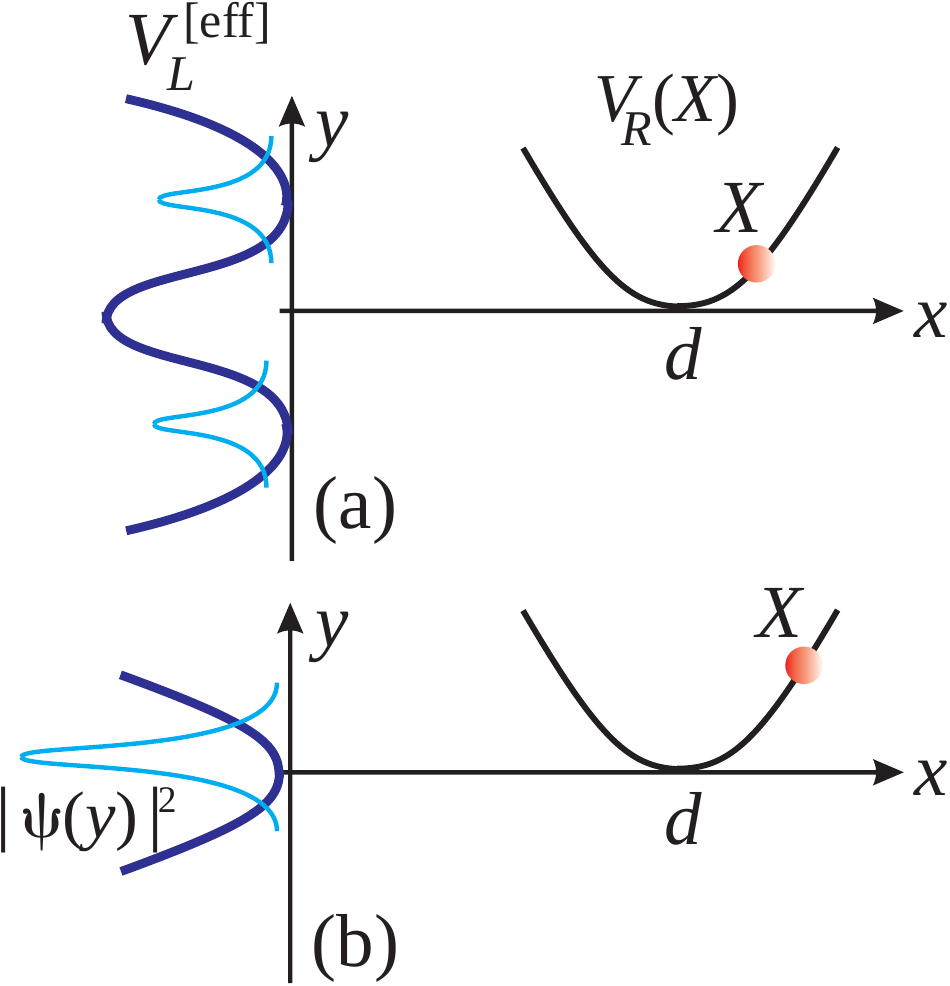}
\caption{Schematic of the two-ion system in the $x \otimes y$ geometry. The left ion is confined along the $y$ direction and treated quantum mechanically with the wavefunction $\psi(y)$ corresponding to the effective potential $V_{L}^{\rm [eff]}.$ The right ion can move along the $x$ axis in the potential $V_{R}(X)$ having the minimum at $X=d$ and acts as an effective classical piston. Panels (a) and (b) illustrate the effective left-ion potential in the double-minimum and single-minimum realization, respectively. Details of the potential and corresponding notations will be presented below in the text.}
\label{fig:system}
\end{figure}

We consider a system of two ions confined in orthogonal, effectively one-dimensional harmonic traps, as illustrated schematically in Fig.~\ref{fig:system}.
The left ion is confined along the $y$ axis in a harmonic potential with frequency $\omega_L$, while its motion in the $x$ direction is frozen by a much stronger confinement. The right ion is confined along the $x$ axis in a harmonic potential with frequency $\omega_R$ and trap center at $d$, while its motion in the $y$ direction is suppressed by tight confinement. 
We assume that the mass of the right ion is much larger than that of the left ion, i.e. the right ion is acting as a piston.
The ions interact via the Coulomb potential, which depends on their relative distance. This geometry allows one to influence the motion of the right ion through the state of the left one. In particular, a time-dependent trap frequency $\omega_{L}(t)$ modifies the quantum dynamics of the left ion and, through the Coulomb coupling, 
affects the motion of the right ion
and thereby enables the control of the effective piston.

In the following, we first introduce the Hamiltonian of the full two-ion system and then derive the corresponding hybrid quantum–classical approximation.

\subsection{System Hamiltonian}

We start from the full Hamiltonian of the two-ion system. The left ion is confined in the trap $V_L$, while the right ion is confined in the trap $V_R$. The total Hamiltonian is written as
\begin{equation}
\hat H=\hat H_L+\hat H_R+\hat H_{\rm int},
\label{eq:hamq}
\end{equation}
where $\hat H_L$ and $\hat H_R$ describe the left and right ions, respectively, and $\hat H_{\rm int}$ accounts for their Coulomb interaction.

The left trap $V_{L}$ is a one-dimensional harmonic potential along the $y$ direction, centered at $y=0$, with time-dependent frequency $\omega_L(t)$,
\begin{equation}
V_L(y,t)=\frac{1}{2}m\omega_L^2(t)y^2,
\end{equation}
where $m$ is the mass of the left ion. The right trap $V_{R},$  a one-dimensional potential along the $x-$ axis having the minimum at the ion position $X=d,$ will be presented below as $V_{R} (X-d).$

The three contributions to the Hamiltonian are then given by
\begin{eqnarray}
&&\hat H_L(t) = \frac{\hat p^2}{2m}+V_L(\hat y,t)
= \frac{\hat p^2}{2m}+\frac{1}{2}m\omega_L^2(t)\hat y^2,\\
&&\hat H_R =  \frac{\hat P^2}{2M}+V_R(\hat X-d),\\
&&\hat H_{\rm int} =V(\hat y,\hat X)
= \frac{C}{\sqrt{\hat y^2+\hat X^2}},
\end{eqnarray}
with $C={q_L q_R}/{4\pi\epsilon_0}.$
Here $q_L$ and $q_R$ are the charges of the left and right ions. In the following, we take $q_L=q_R=e>0$, corresponding to singly ionized ions. The operators $\hat y$ and $\hat p$ denote the position and momentum of the left ion, while $\hat X$ and $\hat P$ denote the corresponding operators for the right ion. 

Let $\Psi(X,y,t)$ denote the wavefunction of the two-ion system. The Schr\"odinger equation can then be written as
\begin{eqnarray}
i\hbar \frac{\partial}{\partial t}\Psi(X,y,t)=\hat H\,\Psi(X,y,t),
\label{eq:SE_full}
\end{eqnarray}
where the total Hamiltonian $\hat H$ is given by Eq.~\eqref{eq:hamq}. In explicit form, this equation reads
\begin{eqnarray}
&&i\hbar \frac{\partial}{\partial t}\Psi(X,y,t)=
-\frac{\hbar^2}{2}\left[
\frac{1}{m}\frac{\partial^2}{\partial y^2}
+\frac{1}{M}\frac{\partial^2}{\partial X^2}\right]\Psi(X,y,t) \nonumber \\
&&+\left[V_L^{[\rm{eff}]}(y,X,t)
+V_R(X-d)
\right]
\Psi(X,y,t),
\label{eq:SE_full_explicit}
\end{eqnarray}
where we have introduced the effective left potential
\begin{equation}
    V_L^{[\rm{eff}]}(y,X,t) = \frac{m}{2}\omega_L^2(t)y^2 +\frac{C}{\sqrt{y^2+X^2}}.\label{eq:Veff}
\end{equation}
This provides the full quantum-mechanical description of the coupled two-ion system, which will be used below as a reference for comparison with the hybrid quantum--classical approximation.


\subsection{Approximation as a quantum-classical hybrid system}

We now derive an approximate description in which the left ion remains quantum mechanical, while the right ion is represented by its classical position and momentum. This approximation is motivated by the large mass imbalance between the two ions and by the fact that the right ion remains strongly localized around its mean position. 
Applying Ehrenfest's theorem for the system, we have the instantaneous change in the expectation value of the position $\hat{X}$:
\begin{equation}
\label{eq:dXdt}
\frac{d}{dt} \langle \hat{X} \rangle = \frac{1}{i\hbar}\langle [\hat{X}, \hat H] \rangle= \frac{1}{i\hbar}\langle [\hat{X}, \hat H_{R}] \rangle=\frac{\langle \hat{P} \rangle}{M}.
\end{equation}
Similarly, we have 
\begin{equation}
    \frac{d}{dt} \langle \hat P  \rangle
    = - \Big\langle \frac{d V_R}{dX} (\hat X-d )\Big\rangle +\frac{1}{i\hbar}\langle [\hat P , \hat H_{\rm int}] \rangle.
\end{equation}
Let $\XX(t) = \langle \hat{X}\rangle$ and $\PP(t) = \langle \hat{P}\rangle$.
We then do the variable change $\hat\zeta = \hat{X} - \XX(t)$ with $\langle \hat{\zeta}\rangle=0.$ {We have also $\langle (\hat X - d )^n \rangle = (\XX(t) - d )^n + \Big\langle\mathcal{O}(\hat \zeta^{2}) \Big\rangle$. Therefore, for $V_R (\hat X - d ) = \sum_{n=2}^\infty V_{R}^{(n)}(0) (\hat X - d )^{n}/{n!}$ with $ V_{R}^{(n)}(0)$ being the $n$th order derivative at zero argument,} we get
\begin{equation}
    \frac{d}{dt} \langle \hat P  \rangle
    = - \frac{d V_R}{dX} (\XX(t) - d ) +\frac{1}{i\hbar}\langle [\hat P , \hat H_{\rm int}] \rangle + \Big\langle\mathcal{O}(\hat \zeta^{2}) \Big\rangle.\label{eq:dpdt1}
\end{equation}

Assuming that $\hat\zeta$ is small, we expand the interaction Hamiltonian around the mean position $\XX(t)$,
\begin{equation}
\label{eq:Hint}
    \hat H_{\rm int}=\hat V (\hat y, \hat{\zeta} + \XX (t)) \approx V(\hat{y}, \XX(t)) + \hat\zeta \frac{\partial V}{\partial X}(\hat{y}, \XX(t))
 + {\mathcal O} (\hat \zeta^{2}).
\end{equation}
Using Eq.\eqref{eq:Hint}, we get
\begin{eqnarray}
 [\hat P , \hat H_{\rm int}] &=& \frac{\partial V}{\partial X}(\hat y, \XX(t))\left[\hat P ,\hat \zeta\right]+\left[\hat P , \mathcal O(\hat \zeta^{2})\right]  \nonumber \\
 &=&-i\hbar\frac{\partial V}{\partial X}(\hat y, \XX(t))+\mathcal{O}(\hat \zeta^2).\label{eq:comPHint}
\end{eqnarray}
Substituting Eq. (\ref{eq:comPHint}) into Eq. (\ref{eq:dpdt1}), we find
\begin{equation}
\label{eq:dpdtnew}
    \frac{d}{dt} \langle \hat P  \rangle =
    - \frac{d V_R}{dX} (\XX(t) - d ) - \Big\langle \frac{\partial V(\hat y, \XX(t))}{\partial X}\Big\rangle+\Big\langle\mathcal{O}(\hat \zeta) \Big\rangle.
\end{equation}
Then the interaction force can be evaluated from the quantum state of the left ion alone,
\begin{eqnarray}
& &\Big\langle \frac{\partial V(\hat y, \XX(t))}{\partial X}\Big\rangle 
= \Big\langle {\Psi}\Big|\frac{\partial V(\hat y, \XX(t))}{\partial X} \Big|\Psi\Big\rangle\nonumber\\
&=& \Big\langle {\psi}\Big|\frac{\partial V(\hat y, \XX(t))}{\partial X} \Big| \psi \Big\rangle +\Big\langle\mathcal{O}(\hat \zeta) \Big\rangle
\end{eqnarray}
where ${\psi} (y)$ normalized with $\int^{\infty}_{-\infty} dy\vert \psi(y) \vert^{2}=1$ is a solution for the left ion alone with the left harmonic oscillator and $H_{\rm int} (t) = V(\hat y, \XX(t))$. 
In detail, $\psi (y,t)$ is a solution of the following time-dependent Schrödinger equation
\begin{eqnarray}
 i\hbar \frac{\partial\psi}{\partial t} = \hat H_{a}(\XX(t),t)\, \psi
 \label{eq:LSE}
\end{eqnarray}
with 
\begin{equation}
\hat H_{a}(\XX(t) ,t) = -\frac{\hbar^{2}}{2m}\frac{\partial^{2}}{\partial y^{2}}
 +\frac{m}{2}\omega_{L}^{2}(t)y^{2}
 + \frac{C}{\sqrt{y^{2}+\XX^{2}(t)}},
\label{eq:Hat}
\end{equation}
meaning that the left ion is located in the potential $V_{L}^{\rm{[eff]}}(y,\XX)$ in Eq. \eqref{eq:Veff} (see Fig. \ref{fig:system}). 

The force acting on the right ion is determined by the wavefunction $\psi(y,t)$ of the left ion and is given by
\begin{eqnarray}
   F(t) &=& -\Big\langle {\psi(y,t)}\Big| \frac{\partial V}{\partial X}\left(\hat{y}, \XX(t)\right)\Big| \psi(y,t) \Big\rangle \nonumber\\
   &=&-\int^{+\infty}_{-\infty} dy \; \vert \psi(y,t)\vert^{2} \frac{\partial V(y, \XX(t))}{\partial X} \nonumber\\\
   &=& C \int^{+\infty}_{-\infty} dy \;  \frac{\XX(t)}{\left(y^{2} + \XX^{2}(t)\right)^{3/2}} \vert\psi (y,t)\vert^{2} .
   \label{eq:pressure}
\end{eqnarray}

Therefore, with the same accuracy, the system of the left trapped ion and its interaction can be described by the Schrödinger equation
\eqref{eq:LSE}.

In the following, we neglect terms $\langle\mathcal{O}(\hat\zeta)\rangle$ and the equations of motion for the right ion shown in Eq. \eqref{eq:dpdtnew} and Eq. \eqref{eq:pressure} can be combined to yield 
\begin{eqnarray}
\label{eq:newton1}
&&\frac{d\XX(t)}{dt}= \frac{\PP (t)}{M},\\
&&\frac{d\PP(t)}{dt}= - \frac{d V_R}{dX} (\XX(t) - d ) +F(t).
    \label{eq:newton2gen}
\end{eqnarray}
These equations show that the position of right ion $\XX(t)$ depends on the left trapped ion with $F(t)$ being given by Eq.~\eqref{eq:pressure}. Thus the system reduces to a self-consistent hybrid quantum–classical dynamics described by the Eqs. \eqref{eq:LSE},\eqref{eq:newton1} and \eqref{eq:newton2}.

For the special case of $V_R$ being an harmonic trap, the second
equations simplifies to
\begin{eqnarray}
&&\frac{d\PP(t)}{dt}=-M \omega_R^{2}(\XX(t)-d ) +F(t).
    \label{eq:newton2}
\end{eqnarray}

It is interesting to note that these equations can be also seen as originating from the combined Hamiltonian/Hamilton function
operator:
\begin{eqnarray}
\hat \cH (\PP,\XX,t) = H_R (\PP,\XX) + \hat H_a (\XX,t)
\end{eqnarray}
where
\begin{eqnarray}
H_R (\PP , \XX ) &=& \frac{\PP ^{2}}{2M}+ V_R (\XX -d ) = \frac{P ^{2}}{2M}+\frac{1}{2}M\omega_{R}^{2}(\XX -d )^{2}\label{eq:HR} \nonumber\\
\end{eqnarray}
is the classical Hamilton function of the right ion.
The time evolution in this quantum-classical hybrid system is then given by:
\begin{eqnarray}
 i\hbar \frac{\partial\psi (t)}{\partial t} &=& \hat {\cH} (\PP (t), \XX(t),t)\, \psi(t) \nonumber\\
 &=& \hat H_a (\XX(t), t) \psi(t) + H_R (\PP(t) , \XX(t) ) \psi(t),\\
\frac{d \XX (t)}{d t} &=& \frac{\partial}{\partial P} \braket{\psi(t)}{\hat\cH(\PP (t), \XX(t),t) \vert \psi(t)} = \frac{\PP(t)}{M}, \\
\frac{d\PP (t)}{d t} &=& -\frac{\partial}{\partial X} \braket{\psi(t)}{\hat\cH(\PP(t), \XX(t),t) \vert \psi(t)} \nonumber \\
&=& -M \omega_R^2(\XX(t)-d)+F(t).
\end{eqnarray}
Since $H_R(\PP,\XX)$ in Eq. \eqref{eq:HR} acts only as a scalar on the wavefunction $\psi$, it contributes only a time-dependent global phase and does not affect observables.


\section{Stationary states \label{sec:stationary}}

In this section we analyze the stationary states of the two-ion system. 
We first present the numerical results and then develop analytical descriptions of the different stationary realizations.

\subsection{Numerical results}

We begin by examining the stationary ground states of the system for different parameters and compare the full quantum treatment with the hybrid quantum-classical approximation.
In the full quantum description, the ground state is obtained from the Hamiltonian $\hat H$ in Eq. \eqref{eq:hamq}. For the hybrid system, the stationary solution of Eqs. \eqref{eq:LSE},\eqref{eq:newton1} and \eqref{eq:newton2} satisfy
\begin{eqnarray}
&& \epsilon_{0} \psi_{0}(y) = \hat H_a (\XX_{0},0)\, \psi_{0}(y),\\
&& \PP_{0} = 0,\\
&&\frac{\partial}{\partial X } \left( H_R (\XX_{0},0) + \epsilon_{0} (\XX_{0})\right) = 0
\end{eqnarray}
with total energy
\begin{eqnarray}
\cE_{0} = H_R (\XX_{0}, 0) + \epsilon_{0}(\XX_0).
\end{eqnarray}

\begin{figure}[t]
\begin{minipage}{\linewidth}
  \centering
  \makebox[0pt][l]{\raisebox{18.5ex}{\hspace{11mm} \large(a)}}
  \includegraphics[width=0.8\linewidth]{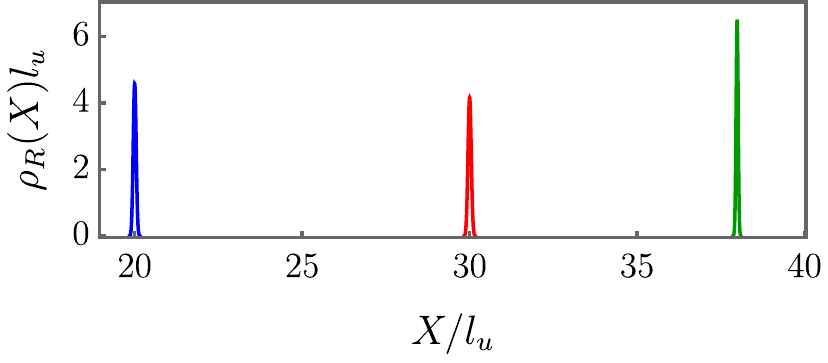}
\end{minipage}
\vspace{1ex} 
\begin{minipage}{\linewidth}
  \centering
  \makebox[0pt][l]{\raisebox{30.0ex}{\hspace{11mm} \large(b)}}
\includegraphics[width=0.8\linewidth]{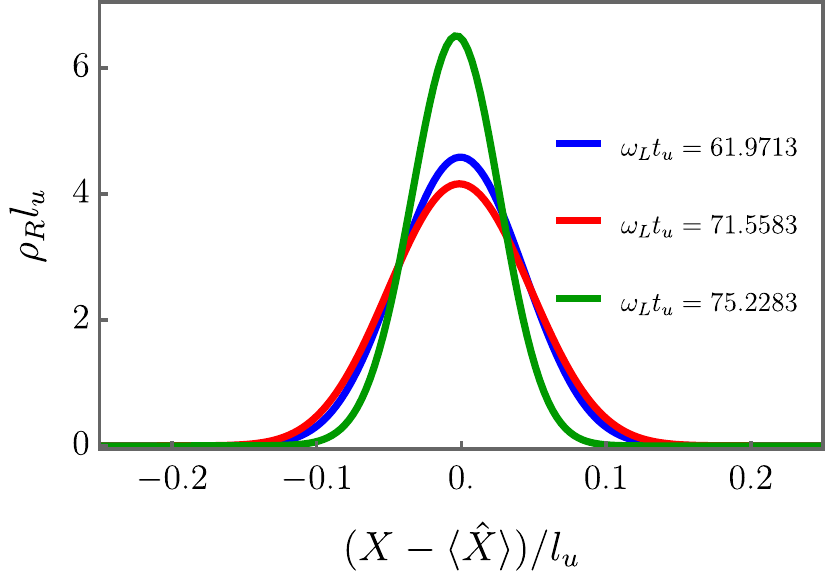}
\end{minipage}
\vspace{1ex}
\begin{minipage}{\linewidth}
  \centering
  \makebox[0pt][l]{\raisebox{19.0ex}{\hspace{13mm} \large(c)}}
\includegraphics[width=0.85\linewidth]{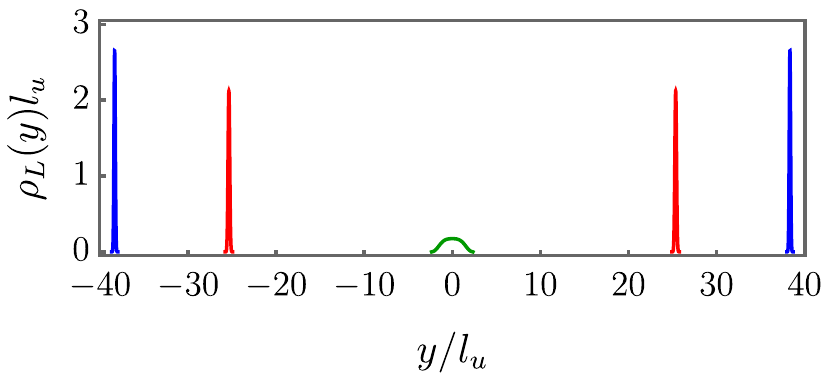}
\end{minipage}
\caption{Stationary state, full quantum treatment: (a) right ion probability densities $\rho_R(X),$ (b) zoomed-in and shifted probability densities $\rho_R,$ and (c) the corresponding left ion probability densities $\rho_L (y)$ for the values of $\omega_{L}$ shown in the plots, see also Eq. \eqref{rhoRL}. Here $\omega_{R}t_u = 20$, $d/l_u=10$.}
\label{fig:fullquantum}
\end{figure}

\begin{figure}[htbp]
\begin{minipage}{\linewidth}
  \centering
  \makebox[0pt][l]{\raisebox{29.0ex}{\hspace{11mm} \large(a)}}
\includegraphics[width=0.8\linewidth]{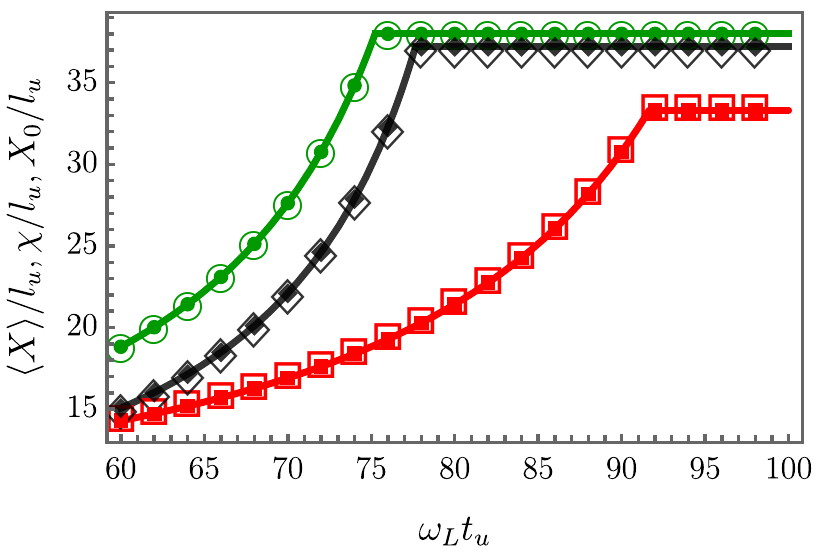}
\end{minipage}\\
\begin{minipage}{\linewidth}
  \centering
  \makebox[0pt][l]{\raisebox{10.0ex}{\hspace{11mm} \large(b)}}
\includegraphics[width=0.8\linewidth]{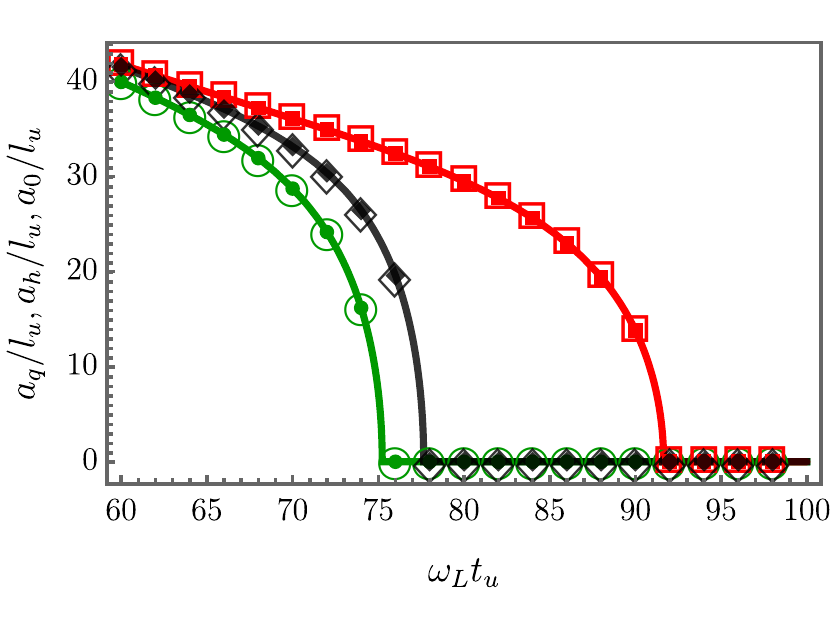}
\end{minipage}
\caption{
Stationary state: (a) Comparisons of positions of right ion among full quantum treatment $\langle \hat X \rangle$ (filled markers), hybrid approximation $\XX$ (open markers) and classical approximation $X_0$ (solid lines) results are shown. (b) Corresponding positions of the left ion in the same three cases: $a_{q},a_{h}$ and $a_0$ parameters as a function of $\omega_{L}.$
Three parameter cases in both figures: $\omega_{R}t_u=20$, $d/l_u=10$ (green), $\omega_{R}t_u=20$, $d/l_u=8$ (black), and $\omega_{R} t_u=25$, $d/l_u=10$ (red). 
\label{fig: 3ab}}
\end{figure}

\begin{figure}[t] 
\includegraphics[width=0.85\linewidth]{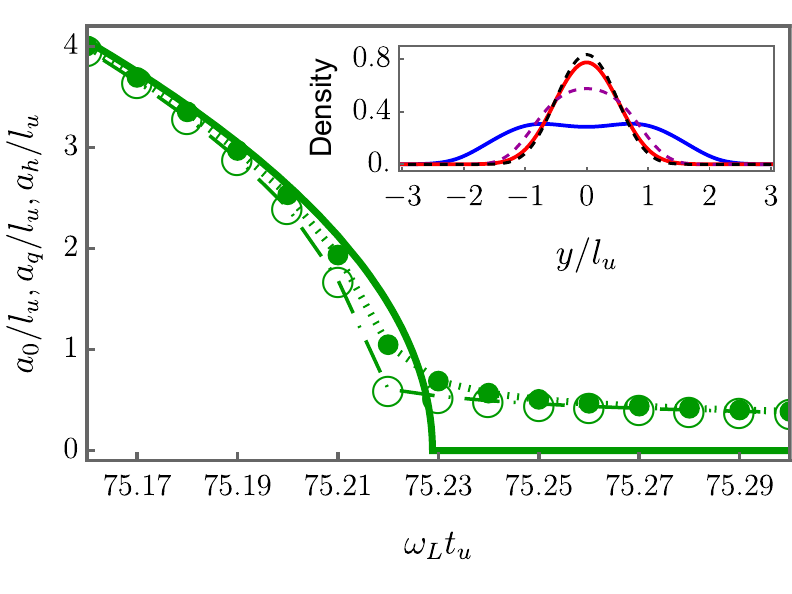}
\vspace{-4mm}
\caption{Stationary state: Zoom in the transition region between the double-minimum and single-minimum regimes of Fig. \ref{fig: 3ab} (b); full quantum treatment (joined filled marks), hybrid approximation (joined open marks), classical approximation (solid line).  Inset: probability density of the left ion in the full quantum calculation $\rho_L(y)$ (solid) and the hybrid approximation $\psi(y)$ (dashed) for $\omega_L t_u=75.22$ (blue solid/magenta dashed) and $\omega_L t_u=75.25$ (red solid/black dashed).}
\label{fig: 4ab}
\end{figure}

In the numerical calculations, we take the left ion to be Beryllium ($\rm{Be}$) with $m=9.012$ a.m.u. and the right ion to be Ytterbium ($\rm{Yb}$) with $M=19.202m=173.05$ a.m.u. {We use the left-ion mass $m$ as the mass unit, length unit $l_{u}=1$ $\mu$m, energy unit $E_{u} = {\hbar^{2}}/ml_{u}^{2}= 7.432\times10^{-31}$ J, time unit $t_{u}=\hbar/E_{u}={m l_{u}^2}/{\hbar} = 0.142$ ms, $1/t_{u} = 2\pi\times 1.122 $ kHz.

We start with the stationary states $\Psi_0(X,y)$ of the full quantum model. Figure \ref{fig:fullquantum} shows the probability densities of the right and the left ion:
\begin{equation}
\rho_R(X)=\int_{-\infty}^\infty dy\, \fabsq{\Psi_{0}(X,y)},\, \rho_L(y)=\int_{-\infty}^\infty dX\, \fabsq{\Psi_{0}(X,y)}, 
\label{rhoRL}
\end{equation}
for different values of $\omega_{L}$.
We can see that the probability densities $\rho_{R}(X)$ in Fig.  \ref{fig:fullquantum}(a) are narrow peaks around the expectation value
\begin{equation}\label{eq:Xexp}
\langle \hat X\rangle =\int_{-\infty}^\infty  \int_{-\infty}^{\infty}  dX dy X\, \fabsq{\Psi_{0}(X,y)}. 
\end{equation}
For clarity, Fig. \ref{fig:fullquantum}(b) shows zooms on the shifted states $\rho_{R}(X - \langle X \rangle)$, illustrating how the width of right ion state changes with $\omega_{L}$. Fig. \ref{fig:fullquantum}(c) shows the left-ion densities $\rho_{L}(y).$ These densities evolve from a two-peak structure to a single-peak one as $\omega_L$ increases, this transition will be discussed in more details below. 

Figure \ref{fig: 3ab} (a) shows the expectation value $\langle X \rangle$ of the right "piston" ion as a function of $\omega_{L}$ (filled marks) and for different values of $\omega_R$ and $d$. 
It is important that one sees two different regimes: $\langle X \rangle$ first increases with the increase in $\omega_{L}$ and becomes a constant after a certain $\omega_{L}.$
Fig. \ref{fig: 3ab} (b) presents the value characterizing the corresponding position of the left ion (filled marks) as: 
\begin{equation}
a_{q} = \left[\int_{-\infty}^\infty dy\, y^{2} \rho_{L}(y)\right]^{1/2}\,.
\end{equation}
We also see two regimes: $a_q$ decreases with the increase in $\omega_L$ and becomes nearly constant after a certain $\omega_L$.

Now we consider the above derived hybrid quantum-classical approximation. The corresponding position of the right ion $\XX_0$ is also shown in Fig. \ref{fig: 3ab} (a). For the left-ion position in the hybrid approximation,
we use a similar definition:}
\begin{equation}
a_{h} = \left[\int_{-\infty}^{\infty} dy \, y^{2} \left|\psi_{0}(y)\right|^{2}\right]^{1/2}.   \label{a_h}
\end{equation}
We again a very good agreement between the full quantum model and the hybrid quantum-classical approximation. 

In the following subsections we derive a further approximation for these two regimes while also approximating both ions in a classical or classically-motivated way and see the details of the quantum transition between regimes presented in Fig. \ref{fig: 3ab}.


\subsection{"Classical" regime with two left-ion peaks}

{If the trap frequency of the left ion is sufficiently weak with $\omega_{L}^{2}<C/mX^{3},$ (see Fig. \ref{fig:system}(a)) 
we approximate the system as an effective classical three-particle configuration, in which the split probability density of the left ion is approximated by two identical point-like peaks located symmetrically at $\pm a_0$.
The Coulomb force acting on the right ion is}
\begin{eqnarray}\label{eq:FC}
F_{\text{C}} = \frac{C X}{(X^{2} + a_0^{2})^{3/2}},
\end{eqnarray}
and the force from the harmonic right trap is
\begin{eqnarray}\label{eq:Ftr}
F_{\text{tr}} = -M\omega_{R}^{2}(X - d).
\end{eqnarray}
Assuming the total system is in stationary, we derive the following conditions:
\begin{eqnarray}
&&\frac{1}{2}\frac{C a_0}{(X^{2} + a_0^{2})^{3/2}} - \frac{1}{2}m\omega_{L}^{2} a_0 = 0, \label{eqs1}\\
&&\frac{C X}{(X^{2} + a_0^{2})^{3/2}} - M\omega_{R}^{2}(X - d ) = 0. \label{eqs2}
\end{eqnarray}
{The factor $1/2$ in both terms of Eq. \eqref{eqs1} is due to the fact that each "part" of the left ion  has half of the charge and mass of the "whole" particle.}

We obtain from \eqref{eqs1} that $X^{2} + a_0^{2}=l_{L}^{2}$ with $l_{L}=\left(C/m\omega_{L}^{2}\right)^{1/3}$.
Using this equality and \eqref{eqs2}, we see that the stationary position of the right ion is $C-$independent and given by
\begin{eqnarray}
X_{0} = \frac{d}{1-m\omega_{L}^{2}/M\omega_{R}^{2}}.
\label{eq: XR2}
\end{eqnarray}
The corresponding stationary values of $a_{0}$ are given by:
\begin{equation}\label{eq:aomega}
a_0(\omega_{L})=\sqrt{\left(\frac{C}{m\omega_{L}^{2}}\right)^{2/3}-\frac{d^{2}}{\left(1-m\omega_{L}^{2}/M\omega_{R}^{2}\right)^{2}}}.
\end{equation}
These two equations present the stationary state as long as $X_{0} < l_{L}.$ Then they are transformed into the single peak solution presented in the next subsection.


\subsection{"Classical" regime with single left-ion peak}

If the left ion is strongly localized close to the centre $y=0$ at  $\omega_{L}^{2}>C/mX^{3}$  (see Fig. \ref{fig:system}(b)), we have two Coulomb-interacting particles. 
The stationary position of the right ion $X_{0}$ in this realization is then determined by:
\begin{eqnarray}
\frac{C}{X_{0}^{2}} - M\omega_{R}^{2}\left(X_{0} - d\right) = 0. \label{eq:localized}
\end{eqnarray}
We find it by solving the cubic equation \eqref{eq:localized}:
\begin{eqnarray}
X_{0} = \frac{d}{3}\left(\frac{d}{\ell}+\frac{\ell}{d}+1\right)>d, 
\label{eq: XR1}
\end{eqnarray}
with
\begin{equation}
\ell = 
\sqrt[3]{d^{3} + \frac{l_{R}^{3}}{2} \left(27 - 3 \sqrt{81 + \frac{12d^{3}}{l_{R}^{3}}}\right)},
\end{equation}
and $l_{R}=\sqrt[3]{{C}/{M \omega_{R}^{2}}}$. 

Note that the position $X_{0}$ in the single-minimum realization is independent of $\omega_{L}$ since the left ion can be approximated as a point particle at $y=a_0=0.$ For this realization $X_{0}$ approaches $l_{R}$ as $d$ approaches 0.  

{To emphasize the difference between the single and double-minima realizations, we calculate the oscillation frequency of the right ion $\widetilde{\omega}_{R}$ near the stationary. In the double-minima regime  we have $M\widetilde{\omega}_{R}^{2}=M\omega_{R}^{2}-m\omega_{L}^{2},$ equivalent to  $\widetilde{\omega}_{R}={\omega}_{R}\sqrt{d/X_{0}}<{\omega}_{R}.$ In the single-minimum realization we obtain $\widetilde{\omega}_{R}={\omega}_{R}\sqrt{1+2l_{R}^{3}/X_{0}^{3}}>{\omega}_{R}.$ Thus, the oscillation frequency of the right ion rapidly increases in the transition between double and single - minimum realizations. This behaviour of the frequency corresponds to the width of the right ion state presented in Fig. \ref{fig:fullquantum}(b), where the state broadens with $X_{0}$ as $\sqrt{\hbar/\widetilde{\omega}_{R}M}$ in the double minima regime and becomes more narrow in the single minimum realization.}


\subsection{Transition between the two regimes: a quantum interval}

The transition between the two regimes occurs at 
$X_{0}=l_{L},$ corresponding to the frequency $\omega_{L}:$
\begin{equation}
\omega_{L}^{2} = \omega_{R}^{2}\frac{M}{m} \frac{(\ell - d )^{2}}{\ell^{2} + \ell d  + d ^{2}}.
\end{equation}
In the case of relatively strong Coulomb interaction with $l_{R}\gg d $, the transition between the two regimes occurs at $\omega_{L}^{2} \approx {\omega_{R}^{2}}M\left(1-d /l_{R}\right)/m$ with $X_{0}$ shifted from $\approx d $ to $\approx l_{L}\approx l_{R}.$ In the opposite $l_{R}\ll d $ case  the transition occurs at $m \omega_{L}^{2}\approx C/d ^{3}$ and sets $X_{0}$ at approximately $d \left(1+l_{R}^{3}/d ^{3}\right).$

This transition between the single- and double-minimum regimes occurs via a very narrow quantum interval where the width of quantum state of the left particle is of the order of the minimum position $\pm a_{0}.$ In this interval $\Delta l\equiv|X-l_{L}|\ll l_{L}$ where the potential $V_{L}(y,X)=-3m\omega_{L}^{2}y^{2}\left(\Delta l-y^{2}/4l_{L}\right)/2l_{L}$ has minima with $a=\sqrt{2l_{L}\Delta l}.$ Analysis of this potential shows that the width of the quantum interval $\Delta\omega_{L}\sim \omega_{L}\sqrt{\hbar/m\omega_{L}}/l_{L}.$ For $X=l_{L}$ the left particle is located at a small $y$ in a quartic potential $V_{L}(y,X)=3m\omega_{L}^{2}y^{4}/8l_{L}^{2}$ and produces a relatively broad state with $a_{h}^{2}\sim {\hbar/m\omega_{L}}\times\left(l_{L}/\sqrt{\hbar/m\omega_{L}}\right)^{2/3}\gg \hbar/m\omega_{L}.$ At a further frequency increase $\Delta\omega_{L}\gtrsim \left(\hbar/ml_{L}^{2}\right)^{2/3}\omega_{L}^{1/3}$ this quartic potential is transformed into a parabolic one with the frequency $\sim\sqrt{\omega_{L}\Delta\omega_{L}}.$

Fig. \ref{fig: 3ab} shows also the classical approximated position of the right ion $X_0$ and the left ion $a_0$ (solid lines) together with the corresponding result of the full quantum treatment and the hyprid approximation. We see a very good agreement between all approches.

{Numerical results for the transition between these two regimes in the corresponding narrow interval of $\omega_{L}$ are presented in Fig. \ref{fig: 4ab}. Since in this interval it is of interest to match the full quantum model and the hybrid quantum-classical approximation, we  compare $\rho_{L}(y)$ with $\fabsq{\psi_{0}(y)}$ where the resulting broadened peaks in the inset of Fig. \ref{fig: 4ab} show a very good agreement between them. In agreement with the previous paragraph, for certain values of $\omega_{L},$ there are two peaks in the smooth probability density distributions since the corresponding $V_{L}^{[\rm{eff}]}(y,X),$ being the sum of the harmonic and Coulomb potentials, has two close minima with a relatively large tunnelling probability, see Fig. \ref{fig:system}. For larger values of $\omega_{L}$ there is only one peak as the corresponding $V_{L}(y,X)$ has then only a single minimum or two very close shallow ones, see Fig. \ref{fig:system}. A visible difference between $\rho_{L}(y)$ and  $\fabsq{\psi_{0}(y)}$ in the inset of Fig. \ref{fig: 4ab} occurs only in a tiny interval of $\omega_{L}$ due to the effect of correction produced by zero-point vibrations ($\sim \hbar/M\omega_{R}X_{0}^{2}$) of the right ion on the potential for the left one.} 


\section{Control of the right ion via inverse engineering scheme \label{sec:dynamics}}

In this section we consider the time-dependent case with the goal of controlling the right ion through modulation of the trap frequency $\omega_{L}(t)$ of the left ion. 
The initial state is the ground state $\Psi_{0}$ at a given $\omega_{L}(0)$ and the target state is the ground state $\Psi_{T}$ at a given $\omega_{L}(t_{f}),$ both in the full quantum treatment. The goal is to design the $\omega_{L}(t)-$dependence corresponding to a desired displacement of the right ion from the initial position at $t=0$ to the final position at $t=t_{f}.$

The key idea will be to first inverse-engineer $\omega_L(t)$ in the classical approximation and then apply this designed $\omega_L(t)$ to the hybrid approximation and full quantum treatment. 

\begin{figure}[t]
\begin{minipage}{\linewidth}
  \centering
  \makebox[0pt][l]{\raisebox{9ex}{\hspace{11mm} \large(a)}}
\includegraphics[width=0.8\columnwidth]{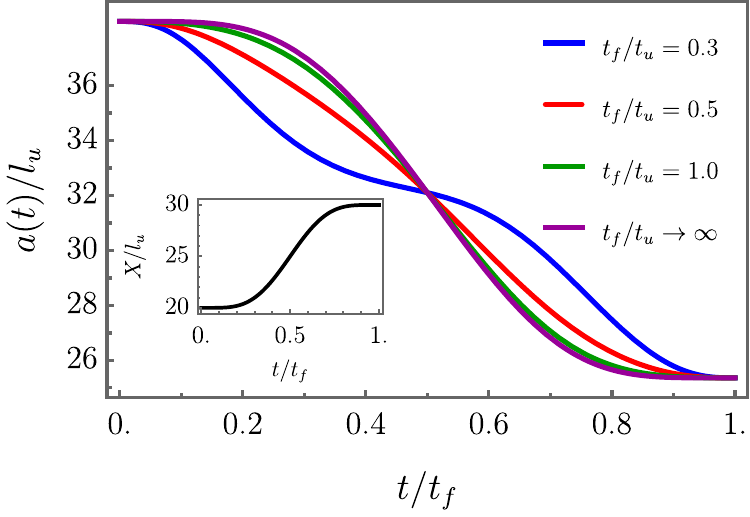}
\end{minipage}
\begin{minipage}{\linewidth}
 \centering
 \makebox[0pt][l]{\raisebox{29.0ex}{\hspace{11mm} \large(b)}}
\includegraphics[width=0.8\columnwidth]{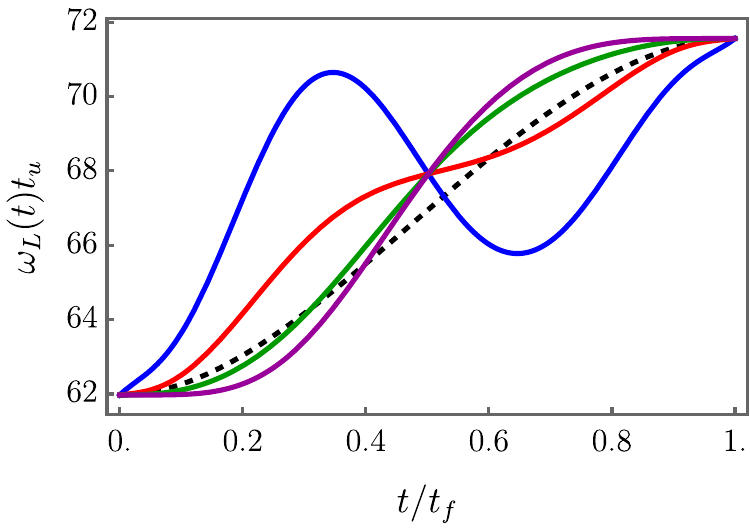}
\end{minipage}\\
\caption{Control schemes corresponding to $X_{0}/l_u=20$ and $X_{f}/l_u=30$, $d_R/l_u=10$, $\omega_R t_u=20$.
(a) Designed position of left particle $a(t)$ for different final times $t_f$. (inset) Designed position of right particle $X(t)$. (b) Designed control schemes $\omega_L(t)$ for different final times $t_f$ as well as the adiabatic reference scheme (dashed line).
\label{fig:control_class}}
\end{figure}

\subsection{Classical inverse-engineered control scheme}

We consider the evolution and control of the two-ion system in the double-minimum regime of the left-ion potential. 
The idea is that we fix first the trajectory of the right ion
$X(t)$ fulfilling the required boundary conditions at initial and final time and then we  inverse-engineer the control function $\omega_L(t)$.

As shown in the previous analysis, the equations of motion for the two ions in this regime are
\begin{eqnarray}
\frac{m}{2} \ddot{a}(t) &=& \frac{C a(t)}{2(X^{2}(t)+a^{2}(t))^{3/2}} - \frac{m}{2} \omega_{L}^{2}(t) a(t), \label{eq:ddy} \\
M \ddot{X}(t) &=& \frac{C X(t)}{(X^{2}(t)+a^{2}(t))^{3/2}} - M \omega_{R}^{2} (X(t) - d ). \label{eq:ddXR}
\end{eqnarray}
Similarly to Eq.~\eqref{eqs1}, the factor $1/2$ in both terms of Eq.~\eqref{eq:ddy} is due to the symmetric probability density distribution of the left ion. Thus, Eq.~\eqref{eq:ddy} is equivalent to the classical equation of motion of a particle of mass $m$ in the potential $V_L(a,X)$. 

From Eq.~\eqref{eq:ddXR}, the trajectory $a(t)$ of the left ion can be expressed in terms of the prescribed right-ion trajectory $X(t)$ as
\begin{equation}
a^{2}(t) = l_{R}^{2} \left(\frac{X(t)}{X(t)-d  + \ddot{X}(t)/\omega_{R}^{2}}\right)^{2/3} - X^{2}(t).
\label{eq:y(t)}
\end{equation}
The time-dependent trap frequency of the left ion then follows from Eq.~\eqref{eq:ddy}:
\begin{equation}
\omega_{L}^{2}(t) = \frac{C}{m \, (X^{2}(t) + a^{2}(t))^{3/2}} - \frac{\ddot{a}(t)}{a(t)}.
\label{eq:omegaLt}
\end{equation}
Thus, the strategy is to first design the trajectory $X(t)$ of the right ion and then determine self-consistently the corresponding functions $a(t)$ using Eq. \eqref{eq:y(t)} and
finally the control function $\omega_L(t)$ using Eq. \eqref{eq:omegaLt}.

We want to start from a given stationary position of the right ion at $t=0$, with $X(0)=X_0$, and end at another stationary position at the final time $t=t_f$, with $X(t_f)=X_f$. Once $X_0$ and $X_f$ are fixed, the corresponding initial and final trap frequencies of the left ion are determined by
\begin{equation}
\omega_{L}(0)=\omega_{R}\sqrt{\left(1-\frac{d}{X_{0}}\right)\frac{M}{m}},
\qquad
\omega_{L}(t_{f})=\omega_{R}\sqrt{\left(1-\frac{d}{X_{f}}\right)\frac{M}{m}}.
\end{equation}

These requirements lead to boundary conditions for the prescribed trajectory $X(t)$. Since the right ion starts and ends in equilibrium, the position, velocity, and acceleration must match their stationary values at $t=0$ and $t=t_f$. In addition, the left-ion coordinate $a(t)$ is not chosen independently, but is obtained from Eq.~\eqref{eq:y(t)} as a function of $X(t)$ and $\ddot X(t)$. Therefore, smooth matching of $a(t)$ to its stationary values at the initial and final times requires
\[
\dot a(0)=\dot a(t_f)=0,\qquad \ddot a(0)=\ddot a(t_f)=0.
\]
where time derivatives of a quantity $a$ will be denoted by $\dot a$.

Differentiating Eq.~\eqref{eq:y(t)} shows that $\dot a$ contains terms proportional to $\dot X$ and $\dddot X$, while $\ddot a$ contains terms proportional to $\ddot X$ and $\ddddot X$ once the lower-order boundary conditions have been imposed. Thus, in order to ensure smooth matching of both $a(t)$ and the derived control protocol $\omega_L(t)$ to their stationary boundary values, we impose
\begin{eqnarray}
&& X(0) = X_{0}, \quad\dot{X}(0) = \ddot{X}(0) = \dddot{X}(0) = \ddddot{X}(0) =0, \nonumber\\
&& X(t_{f}) = X_{f}, \quad \dot{X}(t_{f}) = \ddot{X}(t_{f}) = \dddot{X}(t_{f}) = \ddddot{X}(t_{f}) =0.
\label{eq:BCs}
\end{eqnarray}
These boundary conditions guarantee smooth matching to the initial and final stationary configurations.

Equation~\eqref{eq:y(t)} also implies the existence of a critical final time $t_{f,c}$, which corresponds to the vanishing of the denominator on the right-hand side. For $t_f<t_{f,c}$, the above inverse design is no longer possible. An estimate of this critical time can be obtained as follows. For $X_f>X_0$, one typically has $\ddot{X}<0$ for $t>t_f/2$, with $|\ddot{X}|\sim (X_f-X_0)/(t_f/2)^2$. This yields, for $X_f-X_0\ll X_f-d$,
\begin{equation}
t_{f,c}\sim \frac{2}{\omega_R}\sqrt{\frac{X_f-X_0}{X_f-d}}.
\end{equation}
When $X_f-X_0$ is of the same order as $X_f-d$, this estimate yields
\begin{equation}
t_{f,c}\approx \frac{2}{\omega_R}.
\label{eq:tfc}
\end{equation}

As a reference, we also consider a smooth adiabatic control scheme satisfying $\dot{\omega}_{L}(0)=\dot{\omega}_{L}(t_{f})=0$, namely
\begin{equation}
\omega_{L}^2(t) = \omega_{L}^2(0)+\left(\omega_{L}^2(t_{f}) - \omega_{L}^2(0)\right)(3t_{f}-2t) \frac{t^2}{t_{f}^3}.
\label{eq:omegalref}
\end{equation}

As an example set of parameters, we consider now the initial and final positions $X_0/l_u = 20$ and $X_f/l_u = 30$ and the parameters $d/l_u=10$.
We set also first $\omega_R t_u = 20$.
Figure \ref{fig:control_class} (a) presents the obtained corresponding values for $X(t),a(t),$ and $\omega_{L}(t)$ for different final time $t_{f}$. Note that the chosen function $X(t)$ when plotted as a function of $t/t_f$ becomes independent of $t_f$ by design (see inset in Fig.~\ref{fig:control_class} (a)) while the corresponding function $a(t)$ depends on $t_f$.
The result in the adiabatic limit of $t_f \to \infty$ is also plotted, one can see that here
$a^{2}(t) \to l_{R}^{2} \left({X(t)}/({X(t)-d})\right)^{2/3} - X^{2}(t)$ (where the limit expression depends really only on $t/t_f$). 

The corresponding control schemes $\omega_L(t)$ for this first example are shown in Fig. \ref{fig:control_class} (b). The above introduced reference scheme, see Eq. \eqref{eq:omegalref}, is also shown.
We see that the designed control scheme $\omega_L(t)$ becomes monotonic for large $t_f$, whereas for smaller $t_f$ it develops a minimum and a maximum.
The result in the adiabatic limit of $t_f \to \infty$ is also plotted, one can see that in this limit $\omega_{L}^{2}(t) \to M\omega_{R}^{2} \left(1 - {d}/{X(t)}\right)/m$ (where the limit expression depends again really only on $t/t_f$).


\subsection{Control of quantum state evolution}

While the designed control scheme $\omega_L(t)$ works by construction perfectly in the classical approximation, we now want to examine how good the classically designed control scheme works also in the hybrid quantum-classical description and in the
full quantum treatment.

Here we consider the same example parameters
as above, now for two different values of $\omega_R t_u = 20$ and $\omega_R t_u = 25$. 
Properties of the corresponding stationary states at initial and final time have already been shown in Fig. \ref{fig: 3ab}.
Note that the inverse design of $\omega_L$ breaks down for $t_{f} < 0.12$ in the first case and for $t_{f}<0.1$ in the second one corresponding well to $2/\omega_{R}$ in Eq. \eqref{eq:tfc}.

We start with the hybrid classical-quantum approximation.
As the figure of merit, we examine the relative final classical energy of the right ion
\begin{eqnarray}
Q = \frac{1}{\hbar \widetilde\omega_{R}} \left[\frac{1}{2M} P^{2} (t_f) +  \frac{M}{2}\widetilde\omega_{R}^2 (X (t_f) - X_f)^2\right],
\label{eq:Q}
\end{eqnarray}
where $P(t_f)=M \dot{X}(t)|_{t=t_{f}}$ and as above the effective frequency of the trap of the right ion at final time is
$\widetilde\omega_{R}^2 = {\omega}_{R}^{2} - {m}\omega_L^2(t_f)/M.$ In the classical approximation the designed control scheme for $\omega_{L}(t)$ works by construction perfectly, resulting in $Q=0.$

\begin{figure}[t]
\begin{minipage}{\linewidth}
  \makebox[0pt][l]{\raisebox{11ex}{\hspace{15mm} \large(a)}}
\includegraphics[width=0.85\columnwidth]{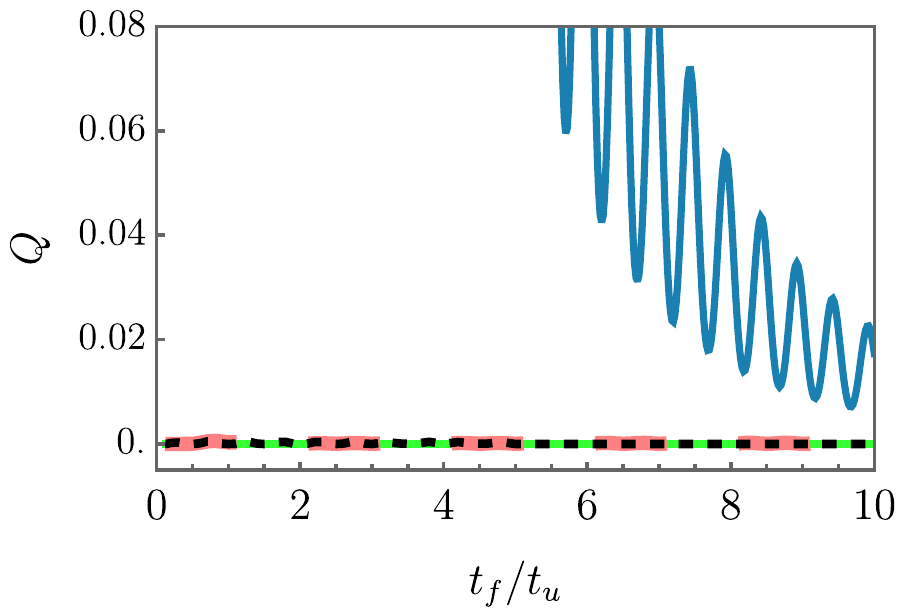}
\includegraphics[width=0.85\columnwidth]{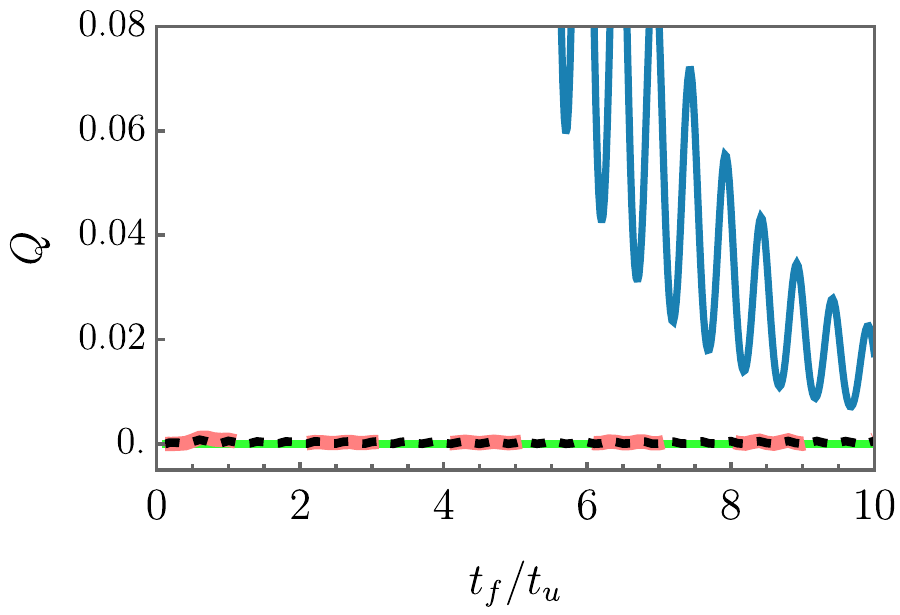}
  \makebox[0pt][l]{\raisebox{11ex}{\hspace{-60mm} \large(b)}}
\includegraphics[width=0.85\columnwidth]{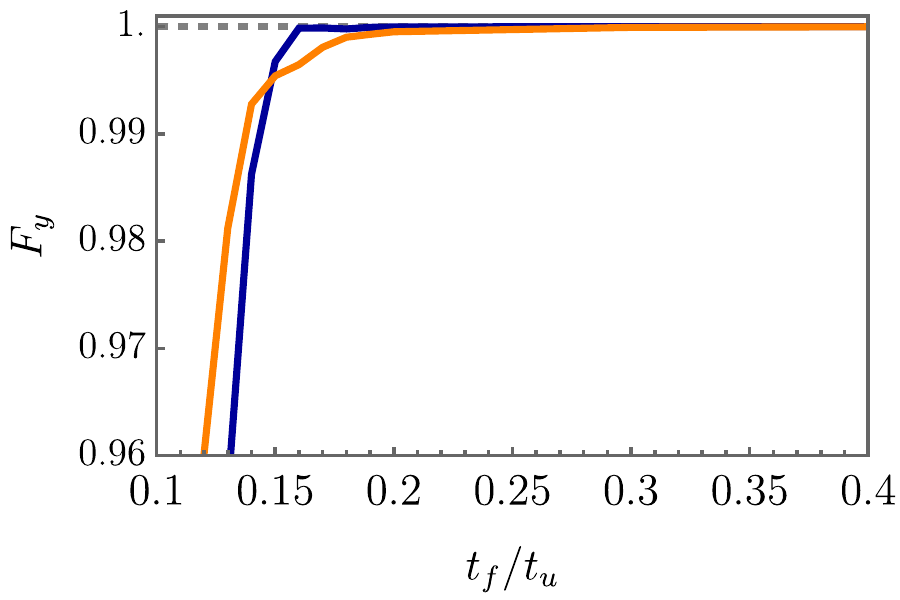}
  \makebox[0pt][l]{\raisebox{10ex}{\hspace{-11mm} \large(c)}}
\end{minipage}
\caption{Performance of the control protocols for the two examples. Panels (a) and (b) show the final excitation measure $Q$ of the right ion, scaled by $\hbar \tilde{\omega}_R$, versus the final time $t_f$ for examples 1 and 2, respectively. The inverse-engineered protocol is shown for the hybrid approximation (dashed red) and the full quantum treatment (solid green). The dotted line indicates the ideal classical result $Q=0$, and the blue curve shows the smooth adiabatic reference protocol. Panel (c) shows the corresponding final-state fidelity $F_y$ of the left ion in the hybrid approximation: $\omega_R t_u = 20$ (orange line) and $\omega_R t_u = 25$ (blue line).}
\label{fig:hybridcontrol}
\end{figure}

The results for the hybrid formalism can be found in Fig. \ref{fig:hybridcontrol} (a) (for $\omega_R t_u = 20$)
and (b) (for $\omega_R t_u = 25$), presented by dashed red lines.
For comparison, the classical $Q=0$ result is also shown by a black, dotted line.
As already mentioned above, the inverse-engineering has a minimal critical time $t_{f,c}$
For $\omega_R t_u = 20$, the critical minimal time is $t_{f,c} = 0.112 \approx 2/\omega_R=0.1$ and
for $\omega_R t_u = 20$, we get $t_{f,c} = 0.089 \approx 2/\omega_R=0.08$. For further comparison, we also show the result obtained by the adiabatic reference control scheme used in the classical formalism (blue solid lines).

We see that the inverse-engineered control scheme works very well not only in the classical framework but especially also in the case of the hybrid quantum-classical framework when the left ion is treated quantum-mechanically.
The adiabatic reference scheme leads to a significant larger final excitation energy $Q$ of the right ion, i.e. it does not work well for the final times shown in these plots.

While the quantity $Q$ does only include the properties of the right, classical ion at final time, we can also quantify how much the wavefunction of the left ion agrees with the corresponding wavefunction in the target stationary state by studying the fidelity
\begin{eqnarray}
    F_y = \fabsq{\int_{-\infty}^{\infty}dy\, \psi_T^* (y) \psi (t_f,y)}
\end{eqnarray}
where $\psi_{T}(y)$ is the wavefunction of the left ion in the stationary target state at $t=t_f$ in the hybrid framework and $\psi (t_f,y)$ is the state obtained after the controlled evolution. For the final times shown this value $F_y$ is close to one in both cases, the values for $t_f/t_u < 0.4$ are shown in  Fig. \ref{fig:hybridcontrol} (c). For $t_f/t_u > 0.2$ we achieve a high fidelity; this means that we also achieve the target wavefunction of the left ion at final time.

We now consider the full quantum treatment. We first consider the value $Q$ which is now defined
by the expectation value of that quantity at the right-hand side of Eq. \eqref{eq:Q} at final time $t_f$ The result is shown as a solid green line in Figs. \ref{fig:hybridcontrol} (a) and (b).
We see a very good agreement between the classical, hybrid and full-quantum framework, i.e. the designed control scheme $\omega_L(t)$ works in all three frameworks.

\begin{figure}[t]
\hspace{2mm}
\vspace{-5mm}
\includegraphics[width=0.9\columnwidth]{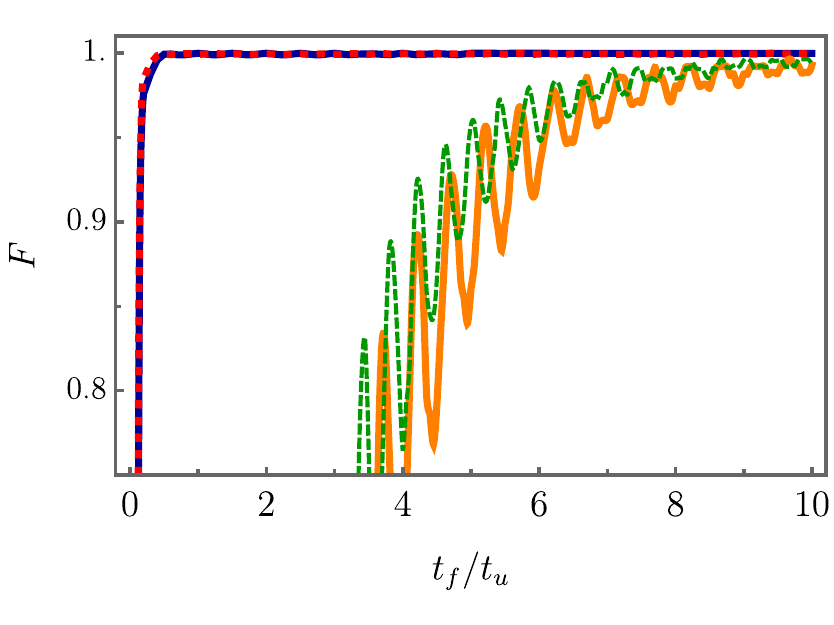}
\caption{Full quantum fidelity versus different final times $t_{f}$ with inverse-designed time-dependent $\omega_{L}(t)$ (first case (solid blue line) and second case (dashed red line)) and a smooth adiabatic scheme (first case (orange solid line) and second case (dashed, green line)).
First case: $\omega_L(0)=61.9713/t_u$, $\omega(t_f)=71.5583/t_u$, $\omega_R=20/t_u$; second case: $\omega_L(0)=77.4641/t_u$, $\omega(t_f)=89.4478/t_u$, $\omega_R=25/t_u$.}
\label{fig:control}
\end{figure}

Another figure of merit in the full quantum framework is the fidelity at final time
\begin{eqnarray}
F = \left| \langle \Psi_{T} | \Psi (t_{f}) \rangle \right|^{2} \, ,
\end{eqnarray}
where $\Psi_{T}$ and $\Psi (t_{f})$ are the exact target ground state and the state obtained by the quantum mechanical evolution, respectively. 
This fidelity is shown in Fig. \ref{fig:control}.
For comparison, the fidelities achieved for the adiabatic control scheme in the quantum-framework is also shown.
For the inverse-engineered control scheme, the fidelity remains close to one for all shown final times $t_f>t_{f,c}$. Here $t_{f,c}$ denotes the critical time below which the inverse engineering breaks down; in both cases we have $t_{f,c}/t_u<0.12$.
On the other hand, the adiabatic reference scheme does not provide a good fidelity until approx. $t_f/t_u > 9$, the final time would need to be larger for the adiabatic scheme to work.

In summary, this shows that the inverse-engineered control scheme results in high fidelities already for very short time in comparison for the times required for the adiabatic reference scheme to work.
Even with the control scheme having constructed based on the classical framework, it also works in the hybrid quantum-classical framework and also in the full quantum framework.

\section{Conclusion}

We have investigated a two-ion system confined in orthogonal harmonic traps, in which the ions are coupled by the Coulomb interaction while moving along perpendicular directions. In this geometry, the lighter ion is treated quantum mechanically and serves as the working medium, whereas the heavier ion can be approximated as a classical particle and acts as an effective piston. This provides a simple but nontrivial setting in which the working medium, the piston, and their mutual backaction can all be described within a single framework.

Using a Be--Yb ion pair as an example, we considered spatial and temporal scales of tens of microns and tens of microseconds, respectively, placing the system in a regime relevant to trapped-ion settings. For the stationary problem, we identified two broad classical regimes connected by a narrow quantum interval. We showed that the hybrid quantum--classical approximation reproduces the full quantum results very well over a wide range of parameters. For the time-dependent problem, we treated the piston control as an inverse-engineering task and derived in this way control protocols that steer the heavy ion between stationary configurations through modulation of the trap frequency of the light ion. These protocols remain effective not only in the classical design framework, but also in the hybrid and full quantum descriptions, where they yield high final-state fidelities and low residual excitation on short time scales.

In this sense, we provided and studied a useful setting in which concepts central to microscopic quantum thermodynamics---such as the identification of a working medium, the emergence of a piston degree of freedom, and their controlled coupling---can be addressed in a simple and physically intuitive way. Future work will focus on the explicit construction of quantum heat-engine cycles in this system, as well as on the role of finite-time operation, energetic cost, dissipation, and fluctuations in such coupled ion-based devices.

\begin{acknowledgements}
J.L. acknowledges that this publication was partly supported by Research Ireland under Grant No. 24/PATH-S/12716.
A.R. acknowledges a research grant from Research Ireland under Grant No. 19/FFP/6951. 
The work of E.S. is financially supported through the Grant PGC2018-101355-B-I00 funded by MCIN/AEI/10.13039/501100011033 
and by ERDF ``A way of making Europe'', and by the Basque Government through Grant No. IT986-16.
\end{acknowledgements}




%

\end{document}